# Photo-electrical properties of 2D Quantum Confined Metal Organic Chalcogenides Nanocrystal Films

*Lorenzo Maserati[†§]\*, Mirko Prato[¶], Stefano Pecorario[†‡], Bianca Passarella[†⊥], Andrea Perinot[†], Anupa Anna Thomas[†‡], Filippo Melloni[†‡], Dario Natali[⊥†] and Mario Caironi[†]\**

[†] Center for Nano Science and Technology @PoliMi, Istituto Italiano di Tecnologia, 20133 Milan, Italy.

[§] The Molecular Foundry, Lawrence Berkeley National Laboratory, Berkeley, CA 94720, USA.

[‡] Department of Energy, Politecnico di Milano, 20156 Milan, Italy.

[⊥] Department of Electronics, Information and Bioengineering, Politecnico di Milano, 20133 Milan, Italy

[¶] Materials Characterization Facility, Istituto Italiano di Tecnologia, 16163 Genova, Italy.

ABSTRACT

2D quantum confined hybrid materials are of great interest from a solid state physics standpoint because of the rich multibody phenomena hosted, their tunability and easy synthesis allowing to create material libraries. In addition, from a technological standpoint, 2D hybrids are promising candidates for efficient, tunable, low cost materials impacting a broad range of optoelectronic devices. Different approaches and materials have therefore been investigated, with the notable example of 2D metal halide hybrid perovskites. Despite the remarkable properties of such materials, the presence of toxic elements like lead are not desirable in applications and their ionic lattices may represent a limiting factor for stability under operating conditions. Alternative, non-ionic 2D materials made of non-toxic elements are therefore desirable. In order to expand the library of possible hybrid quantum wells materials, here we consider an alternative platform based on non-toxic, self-assembled, metal-organic chalcogenides. While the optical properties have been recently explored and some unique excitonic characters highlighted, photo-generation of carriers and their transport in these lamellar inorganic/organic nanostructures, critical optoelectronic aspects, remain totally unexplored. We hereby report the first electrical investigation of the air-stable $[AgSePh]_\infty$ 2D coordination polymer in form of nanocrystal (NC) films readily synthesized *in situ* and at low temperature, compatible with flexible plastic substrates. The wavelength-dependent photo-response of the NC films suggests possible use of this materials as near-UV photodetector. We therefore built a lateral photo-detector, achieving a sensitivity of 0.8 A/W at 370 nm thanks to a photoconduction mechanism, and a cutoff frequency of ~400 Hz, and validated its reliability as air-stable UV detector on flexible substrates.

MAIN TEXT

Bottom-up, atomic-scale strategies for the synthesis of functional materials enable molecular level control of macroscale properties of the overall assembly.[1] Many fruitful examples span the area of metal organic frameworks,[2] conductive coordination polymers,[3] metal-organic chalcogenides.[4] Recently, these technologies have generated notable breakthrough advancements in the optoelectronic field, by introducing tunable self-assembled quantum wells, with the notable example of the two-dimensional (2D) metal halide perovskites.[5,6] These materials offer the ease of handling and environmental stability of solution-processable bulky materials, as opposed to pure 2D materials, joint with quantum properties typical of atomically thin compounds. Despite the investigation of 2D hybrid perovskites set its roots in the XIX century,[7] only about 25 years ago researchers started to consider hybrid multiple quantum wells (HMQW) as new fundamental structures for room-temperature excitonic-based optoelectronics.[8–10] The field mildly developed in the following decade, broadening the exploration to the 2D-like metallorganic chalcogenides.[11,12] Owing to the surge of metal halides perovskites in 2010's,[13] a fresh, renewed interest in HMQW swept the scientific community. In the last five years, research on 2D hybrid perovskites boosted,[14] driven by the promise of facile synthesis, optoelectronic properties tunability and broadband applications in light absorbing or emitting devices.[5–7,15] On the other hand, the intrinsic instability of the ionic lattice to moisture together with other non-idealities occurring under real working conditions, are problematic in 3D perovskites, and also in their 2D analogs, where lead-based materials are still the top-performers.[16] Within this context, alternative material platforms with similar HMQW characteristics that could provide improved intrinsic environmental stability and that at the same time exclude toxic heavy metals are desirable. Very recently, ideas[12,17] and materials[4] developed in applied physics and inorganic chemistry have been successfully employed to lay the foundation for the systematic exploration of covalently bound coordination polymers based on lamellar metal organic chalcogenides (MOC).[18–20] In particular, the silver benzeneselenolate [AgSePh]$_\infty$ was shown to host tightly bound anisotropic excitons featuring sharp adsorption lines and narrow photoluminescence, while offering optoelectronic tunability by substitution of the chalcogen anion.[20] Although the peculiar excitonic properties have been demonstrated, the possibility to photo-generate and transport charge carriers, fundamental aspect of most optoelectronic devices, is still unexplored. Indeed, much of the reported literature on coordination polymers and metal-organic frameworks, with the only exception of 2D perovskites, focused selectively on only one of the two fundamental semiconductor properties: either absorption-photoluminescence,[19] or charge carriers transport[21] (typically investigated in heavily doped systems). In fact, metal organic coordination polymers in which both semiconductor traits have been shown are scarce.

In this article, we investigate the so far unexplored photo-electrical properties of the [AgSePh]$_\infty$ coordination polymer, whose charge carriers were shown to undergo 2D quantum confinement by optical spectroscopy. Our results suggest that MOC platform has the potential to provide self-assembled, low temperature and air-stable materials for UV detection applications. [AgSePh]$_\infty$ nanocrystal (NC) films were prepared according to an optimized synthesis yielding reproducible and air-stable samples. We investigated the photo-response in two terminal, lateral devices, characterized by dark currents with a large thermal activation energy of 1.10 eV. The devices showed a wavelength-dependent photocurrent response, with the spectral shape resembling the [AgSePh]$_\infty$ absorption spectrum up the band-edge at 405 nm. There, a photoconduction mechanism appears to boost the device responsivity up to 0.8 A/W at 370 nm. The lateral device is characterized by a cutoff frequency around 400 Hz and a light-to-dark ratio approaching 1000, for broadband illumination. The versatility of the synthesis allowed us to explore the device fabrication on several substrates and to anticipate an appealing application of the [AgSePh]$_\infty$ NC film in air-stable, flexible UV photodetectors.

RESULTS AND DISCUSSION

Silver benzeneselenolate [AgSePh]$_\infty$ (Ph = phenyl groups) has a monoclinic unit cell (Figure 1a) that gives rise to a layered crystal where the inorganic planes are non-covalently bonded along the [001] direction. The [AgSePh]$_\infty$ NC film shows a characteristic absorption spectrum with 3 excitonic resonances in the blue-violet region (Figure 1b).[20] UV photoemission spectroscopy (UPS) reveals a deep (-5.34 eV) valence band maximum (*VBM*) (Figure 1c,d) with a finite density of states (Figure 1d, inset) approaching the Fermi energy ($E_F$), positioned 1.39 eV above the *VBM*. Considering the previously reported transport bandgap of 3.05 eV,[20] the $E_F$ lies slightly below the half-gap, indicative of possible p-doping. A few electronic states positioned above the *VBM* are suggested to be localized states contributing the absorption tail in the visible spectrum (Figure 1b).

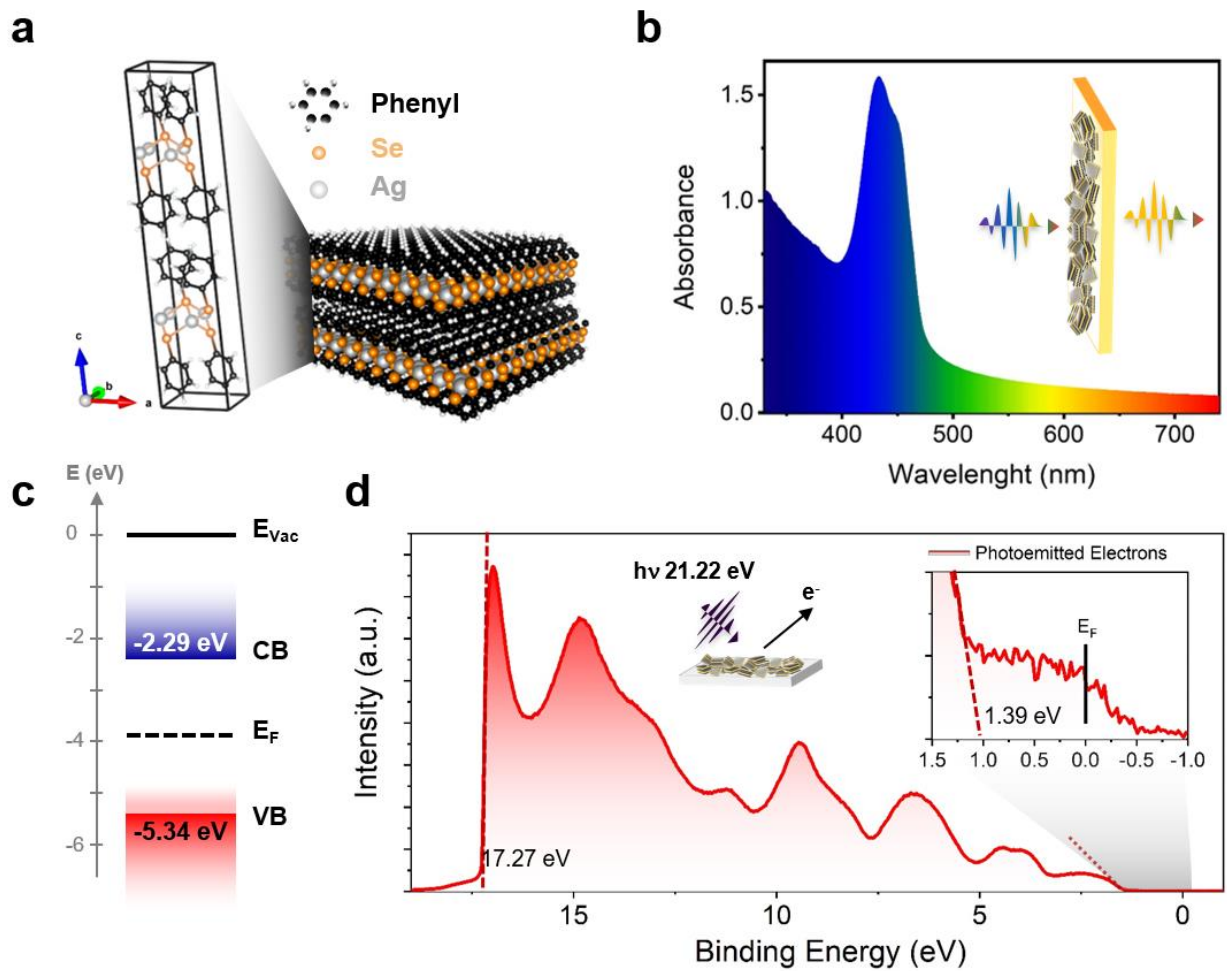

**Figure 1. [AgSePh]∞ optoelectronic characteristics. a,** Unit cell and polymeric crystal structure of the material. **b,** Absorption spectrum in the UV-Vis range showing the excitonic resonance in the blue region. **c,** Energy diagram of [AgSePh]∞ based on the ultra-violet photoemission spectroscopy (UPS) **d,** UPS data collected by irradiating the sample with 21.22 eV photons showing a work-function of 21.22-17.27 eV = 3.95 eV (Fermi level) and a valence band maximum (*VBM*) located at –(3.95+1.39) eV = -5.34 eV. Considering previous experiments reporting the band-gap value of 3.05 eV,[20] the conduction band minimum (*CBM*) is then to be placed at -2.29 eV. Inset, zoom-in showing finite density of states approaching the Fermi edge.

The [AgSePh]$_\infty$ NC film was obtained following a 3-steps process described in detail in a separate work, here schematized in Figure 2a and summarized in the Methods section. Briefly, a substrate of choice was covered by 20 nm of thermally evaporated silver, then mildly oxidized by O$_2$ plasma and subsequently reacted, by a sort of a chemical vapor deposition (CVD), in an inert atmosphere saturated by vapor of benzeneselenol at 90 °C. The reaction left no detectable trace of unreacted crystalline silver nor silver oxide, while the crystal structured obtained by grazing angle X-ray diffraction nicely matches the calculated pattern[22] (Figure 2b). Scanning electron microscopy (SEM) and atomic force microscopy (AFM) reveal that NC composing the film have lateral sizes spanning from ~50 nm to ~300 nm, where dimensional quantum confinement can be excluded. The whole NC film looks very homogeneous over millimeters scale (see SEM images in Figure S1), with an average thickness of ~250 nm, as measured by scanning profilometry.

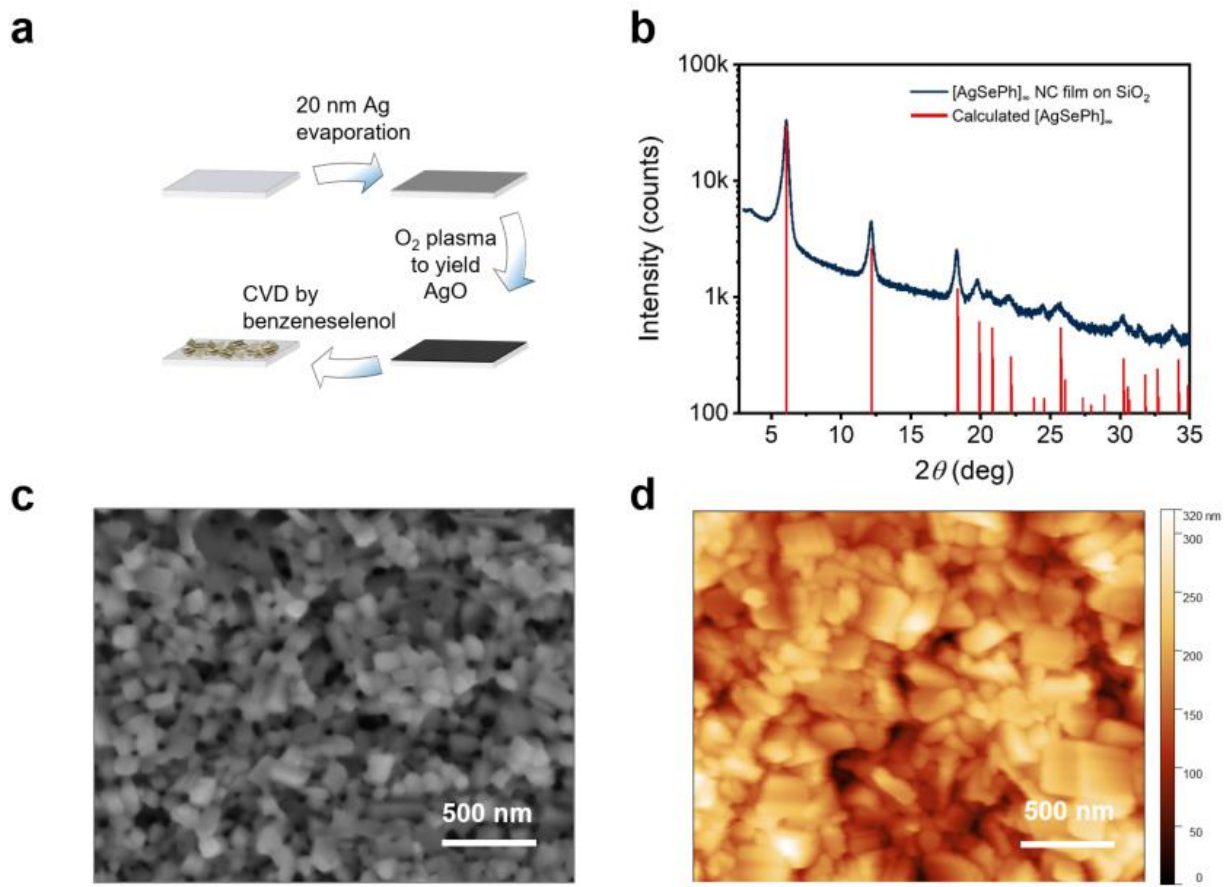

**Figure 2. [AgSePh]∞ NCs films with optimized synthesis, crystallinity and morphology. a,** Synthetic approach to yield [AgSePh]∞ NC thin films. Any substrate of choice can be covered by 20 nm silver, then oxidized by O$_2$ plasma. The atomic oxygen promotes a fast reaction in the benzeneselenol saturated atmosphere fully converting the AgO thin film into a [AgSePh]∞ NCs film **b,** Thin film X-ray diffraction (XRD) data compared to the calculated structure.[22] **c-d,** SEM and AFM images, respectively, of the thin film showing distinct nanocrystals.

The film uniformity was further confirmed on centimeters scale by optical imaging directly on the devices used for electrical characterization (Figures. 3a,b). Several substrates (SiO$_2$/Si$^{++}$, low-alkali glass, polyethylene 2,6-naphthalate (PEN)) with pre-patterned interdigitated electrical contacts were used to test their effect on the NC film growth. No appreciable differences, optical or electrical, were detected, as emerged from the hereafter illustrated results, further demonstrating the synthetic approach reliability and versatility. Figure 3c shows the current vs voltage characteristics of a typical sample of [AgSePh]$_\infty$ NC films deposited on SiO$_2$/Si$^{++}$ with pre-patterned, interdigitated electrodes with an equivalent width of 10 mm and a varying inter electrode length from 5 to 20 µm.

Different metal contacts have been explored for the optimization of the device, including Pt, Au, Al. Dark currents are in all cases very similar, in the 10 pA range for a bias of 5 V. Platinum contacts lead instead to the best results both in terms of reduced hysteresis and of the highest photo-current under ~110 mW/cm$^2$ white light illumination (10 nA at 5 V), while aluminum to the lowest (2 nA at 5 V, see Figure S7). In the following, we refer to devices with Pt contacts.

Not only the dark current is found to increase as channel length ($L$) decreases as expected (see Figure S2b), but also the current under white light conditions $I_{light}$ appears to scale super-linearly with $1/L$ (actually it is proportional to $L^{-1.29}$ at a fixed applied voltage of 5 V, as shown in Figure S2a). The light to dark current ratio reaches ~10$^3$ for 20 µm channels. The dependence of the photocurrent on $L^{-1.29}$ rules out: 1) regimes where the photocurrent is proportional to the illuminated area (and carriers transport does not constitute a bottleneck) because $I_{light}$ would scale as $L$; 2) regimes where the photoactive area is smaller than the inter-electrode spacing, such as in the collection limited and in the space charge limited photocurrent regimes.[23,24] We will come back on the photoconductive mechanism later.

Temperature dependent current vs. voltage (*IV*) curves (Figures S3a,b) were used to extract the currents at a fixed voltage and plot the dark and photo-conductivity over temperature (Figure 3d). From the dark conductivity plot in natural semi-log units versus $1/k_BT$, the thermal activation energy ($E_A$) needed by the charges to be injected and transported across the film according to the Arrhenius equation [$\sigma(T) = \sigma_0 e^{E_A/k_B T}$] can be extracted. Our linear fit reported in Figure S3c shows an $E_A$ of 1.10 eV for the only data points available above the instrumental noise level. This thermal activation energy could indicate a transport mechanism based on charge hopping across NCs and/or defect localization within the single NC. In addition, we cannot exclude contributions from the injection mechanism. Of course, a single crystal study would be desirable to disentangle the contributions, but technical limitations on the crystal size and thickness prevented this investigation so far. On the other hand, the variable temperature photocurrent (see Figure 3d and Arrhenius plot in Supporting Figure S3d) points instead to a more complex behavior where, for sake of comparison, we consider the current data points restricted to the range where the dark current was measurable. This resulted in a much lower $E_A$ of 0.23 eV for the photo-generated charges. At lower temperature the slope of $\sigma(T)$ reduces further, apparently towards a temperature independent regime, a possible fingerprint of transport dominated by tunneling through localized states.

To further weigh in on this analysis we discuss the quality of the [AgSePh]$_\infty$ NC films, assessing the eventual presence and possible effects of unreacted metal reagents in the electrical characterization. The O$_2$ plasma oxidation of metallic Ag thin film produced AgO, which shows a finite density of states at the Fermi energy (UPS, Figure S4). The electrical behavior of AgO depends upon the plasma oxidation extent both in terms of duration and strength: the harsher the plasma conditions, the lower the AgO conductivity (Figure S5a). Nevertheless, the *IV* curves remain linear upon oxidation and show a conductivity ranging from of 6 nS/cm to 20 µS/cm. These values largely exceed the 1 pS/cm of dark conductivity in the [AgSePh]$_\infty$ NC film. Indeed, in case the mildly oxidized Ag used for the optimized CVD reaction were largely present in the reacted film, it would dominate the overall electrical behavior (as it tops the conductivity of [AgSePh]$_\infty$ by 5 orders of magnitudes). We recall here that XRD, X-ray photo-emission spectroscopy (XPS) and optical spectroscopy were not able to detect tangible AgO leftovers. Overall, we can conclude that eventual non-converted AgO following the CVD reaction, if present, are not continuous and do not affect the electrical properties.

We then consider the values of dark and photo-current in [AgSePh]$_\infty$ NC film grown under different conditions as indicators for the CVD reaction completion and we find that: i) mild Ag oxidation is preferable, as it leads to lower dark currents in [AgSePh]$_\infty$ and slightly higher photocurrents (Figure S5b), possibly due to rougher, less continuous silver oxide film following harsh plasma conditions; ii) CVD reaction times longer than 1 h, and up to 24 h, do not lead to statistically significant electrical differences in the NC film (Figure S5c,d); iii) the final film thickness, tunable by changing the thickness of the starting Ag film, does not affect dark and photo current values in the range from 250 nm to 500 nm (Figure S6a,b); effects with thickness can be measured only for very thin, non-continuous layers, as for example those with a nominal thickness of 40 nm.

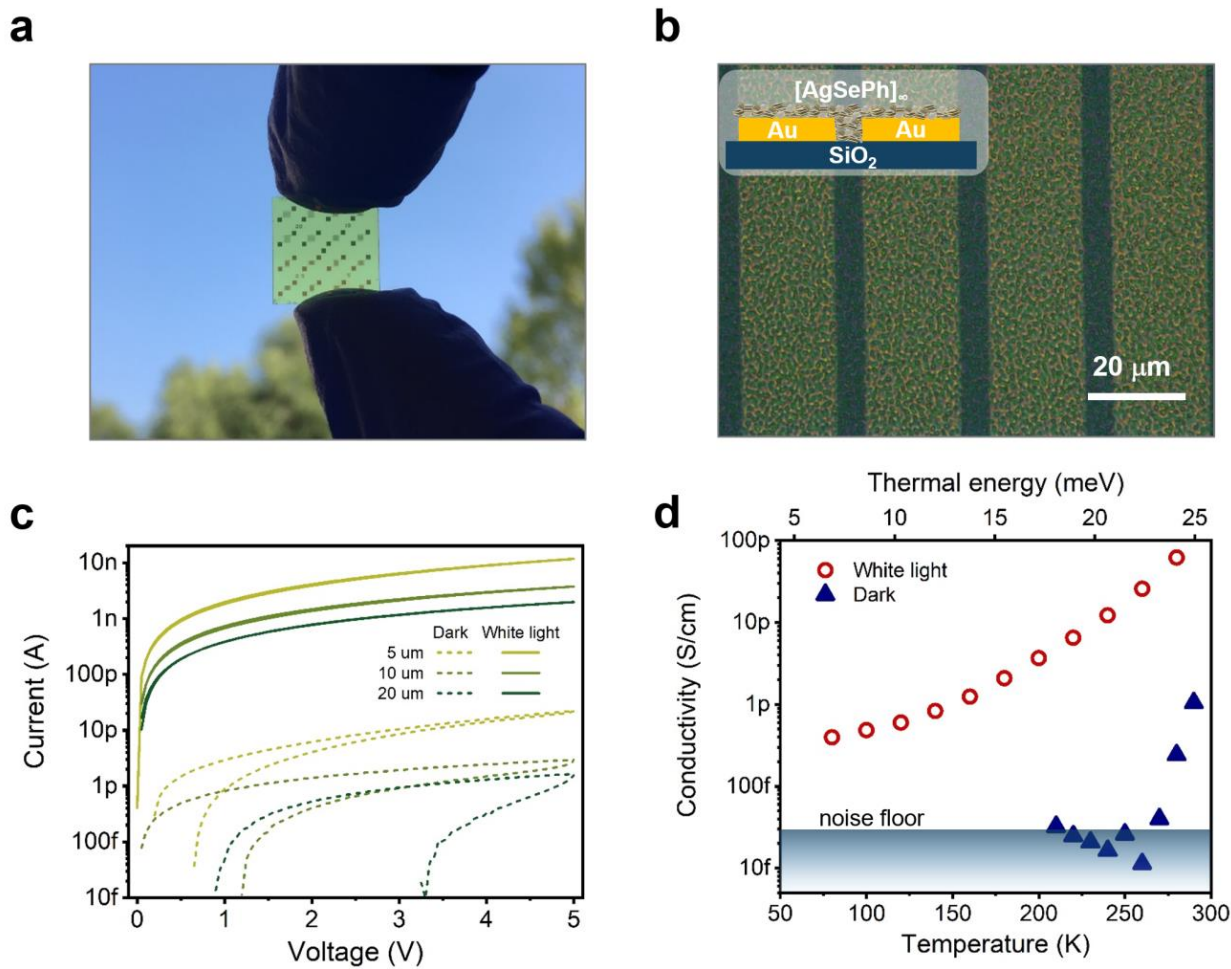

**Figure 3. [AgSePh]$_\infty$ NCs films electrical properties. a.** Photograph of semi-transparent devices used to perform part of the electrical characterization. **b.** Optical microscope image of a single device (inset, device schematic). **c.** IV curves under dark (dotted lines) and white light illumination (~110 mW/cm$^2$, solid lines) for different channel lengths (5, 10, 20 μm, represented in light green, green, dark green, respectively) showing scaling behavior at constant applied voltage. **d.** Current maximum value (5 μm channel length) at fixed potential (5 V) versus temperature. Platinum

contacts were used as they produced the smallest hysteresis in IV curves and the highest photocurrent.

This fact requires a separate explanation for dark and photocurrent. As to the latter, we observe that the absorption length at the peak is about 45 nm, therefore it is reasonable that the absorbed amount of photons does not change appreciably on going from 250 nm to 500 nm. As to the former, since the dark resistance of the device does not change with the active layer thickness, either the contacts are the limit in dark conditions, or a combined effect of a non-uniformity of the material in the vertical direction and of field-assisted transport, with the field peaking in the close proximity of the 40 nm thick electrode, is hiding a sizeable contribution of the top film parts to dark currents.

We then turn the focus on the contributions of different light wavelengths to photo-carriers generation. Several LEDs were used to shine monochromatic light spanning the extended visible range (wavelength $\lambda$ from 370 nm to 810 nm) on a device (see Supporting Figure S8 for raw data and details). We then calculated the device responsivity, defined as $R(\lambda) = (I_{light} - I_{dark})/P = I_{pc}/P$ (where $I_{pc}$ is the net photocurrent and $P$ is the incident optical power). The responsivity spectrum for a device with $L$ = 5 μm at an applied bias of $V$ = 5 V is shown in Figure 4a. The highest photo-response is obtained in the UV range. This is interesting since the maximum responsivity does not match the maximum absorption, peaked on the excitonic resonances. Instead, this photo-response concerns free-carriers generated more efficiently from above-bandgap excitation. Tightly bound excitons on the other hand are more likely to decay without generating free-carriers under the weak electrical fields applied. The device achieves a peak responsivity of ~0.8 A/W at 370 nm. As a comparison, commercial Si photodiodes show $R \approx 0.2$ A/W in the 400 nm wavelength range, dropping by more than an order of magnitude around 300 nm.[25]

The good responsivity of our simple device implies an efficient detection mechanism. A clear indication of the working mechanism comes from the corresponding external quantum efficiency value (EQE), which at 370 nm corresponds to approximately 270 %, implying the presence of a photoconductive gain. Photoconduction sees the presence of a trapped carrier and the possibility to recirculate the mobile one, i.e. to extract and reinject it to produce a gain. Given the use of high work function electrodes and their nominal energy level alignment with respect to the semiconductor bands (Figure 1d), we speculate that the photoconduction mechanism in the $[AgSePh]_\infty$ NC devices is based on hole collection and recirculation, and electron trapping. Such mechanism is compatible with the observation of the highest photocurrent for the highest work function electrode (Pt). In principle, the photocurrent in a lateral photoconductor would scale as $1/L$. The fact that it actually scales with $L^{-1.29}$ can be explained as due to a field dependence for carriers' photo-generation and/or a field-dependence for electron mobility. Such field-dependent photoconduction mechanism has been previously observed in organic and perovskites photoconductors.[26,27]

We then carried out a specific investigation of this device at $\lambda = 370$ nm to better characterize its UV response by measuring *IV* curves under monochromatic light (in Figure 4b with dark *IV* for comparison) and photocurrent over time under intermittent illumination (Figure 4c) at increased intensities. This latter analysis shows a stable photo-detector response over three orders of magnitude of light intensity. Figure 4b shows a nearly symmetric response (approaching the hysteresis-free response at very slow scanning rate) for both dark and photocurrent, typical of photoconductors. To verify if the photo-carriers extraction rate was influenced by an optical filter effect related to the thickness of the NC film, a corning glass substrate was used such that the sample could be illuminated both from the top and from the bottom under the same conditions. No relevant differences given the experimental uncertainties between top and bottom illumination were recorded (see Figure S9).

Figure 4d shows on a bi-logarithmic scale the photocurrent generated by increased optical power per unit of area, where two distinct regimes can be found: i) below 1 µW/cm$^2$ the photocurrent scales almost linearly with the light intensity ($I_{pc} \propto P^{-0.95}$), and the device responsivity is practically constant;[28] ii) above 1 µW/cm$^2$ the slope of the photo-generated current increase progressively slower ($I_{pc} \propto P^{-0.62}$) as a consequence of the reduction of $R$ with light intensity. This latter behavior is compatible with a multiple trapping and release transport model for holes, with an intra-gap DOS given by an exponential distribution. The gain reduction with power intensity is due to the fact that more recombination centers (actually trapped electrons), are present at higher photo-generation rates: the hole lifetime decreases and so does the photoconductive gain.[28,29]

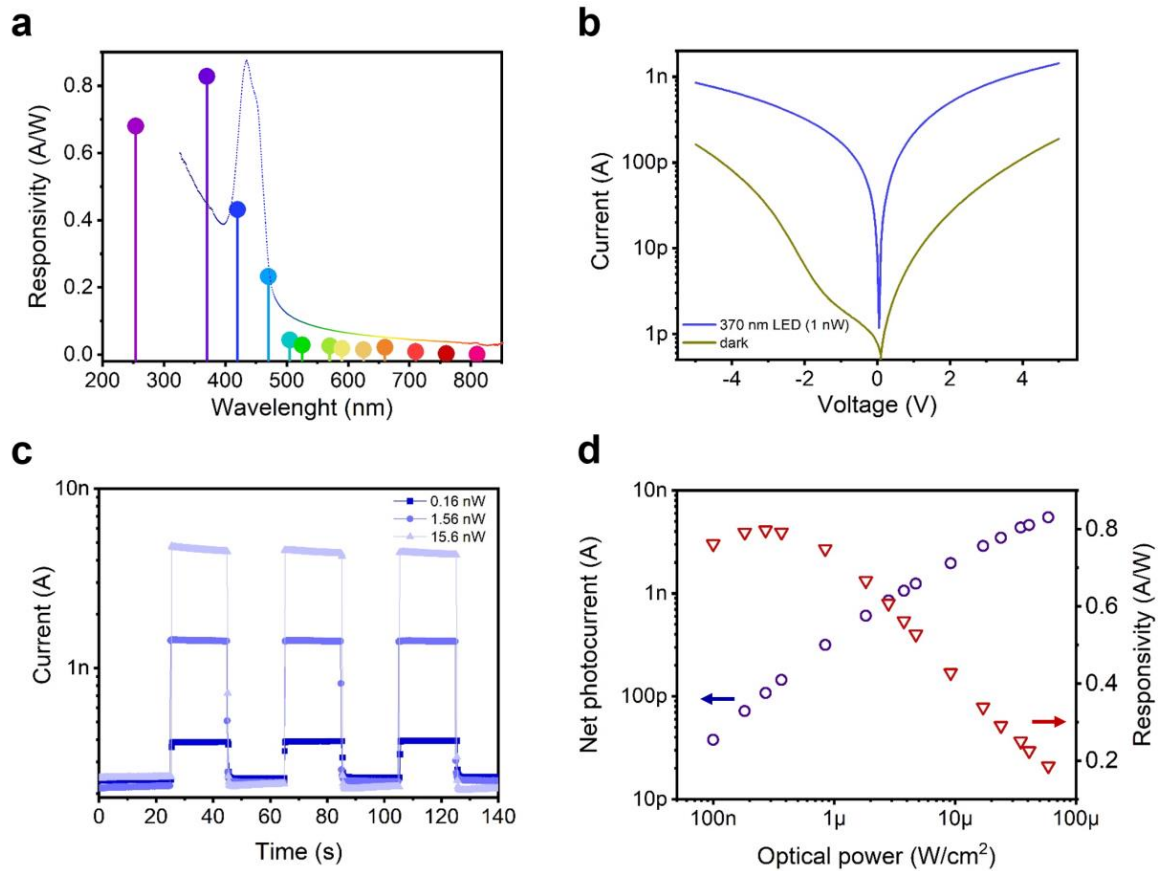

**Figure 4. [AgSePh]∞ planar UV photo-sensor characteristics (platinum contacts, device active area ≃0.05 mm$^2$). a,** [AgSePh]∞ NC film photocurrent response, normalized over incidence optical power, *i.e.* photo-sensor responsivity at different wavelengths (constant irradiance ~300 nW/cm$^2$); the absorption profile of Figure 1b is overlaid as dotted line for reference. **b,** *IV* curves sensor characteristics in dark conditions and under 370 nm monochromatic (LED) illumination (1 nW integrated over the whole device area). **c-d,** Photo-current response at 370 nm and 5 V applied bias. In panel **c**, the light stimulus was repeatedly switched between on and off, and the sensor, tested with three different illumination powers (train of rectangular pulses), shows stable and reproducible time response. In panel **d**, the net photocurrent and the responsivity are plotted as a

function of the incident power density. At 60 µW/cm² the device delivers photo-current as high as 5.5 nA, whereas the maximum responsivity, reached at lower irradiance, is 0.8 A/W.

Given the photo-conductive nature of the photo-response, it is interesting to assess the dynamic behavior of the device, as under such working regime the response time is typically traded-off for responsivity. To this purpose, we measured the device frequency response under oscillating illumination with a custom-made setup (see schematic in supporting Figure S10a and setup control experiment in the Supporting Figures S10b,c). A sinusoidal voltage at varying frequency was generated by a network analyzer and sent along with an offset bias to a UV LED ($\lambda$ = 370 nm) to modulate its optical emission $W_{in}$ over a steady-state illumination. The photocurrent generated by the light modulation was sent back to the network analyzer as voltage signal through a transimpedance amplifier ($V_{out}$). We then plot the ratio of the output/input signals over frequency. The measurement was repeated for different bias voltages applied to the detector (Figure 5a). Considering the measurement noise and the associated uncertainty, the cutoff frequency, defined as the drop of $V_{out}/W_{in}$ 3 dB below the steady state, is in the 300 – 500 Hz range and does not show a sizeable dependence on the detector bias (and so does the phase response).

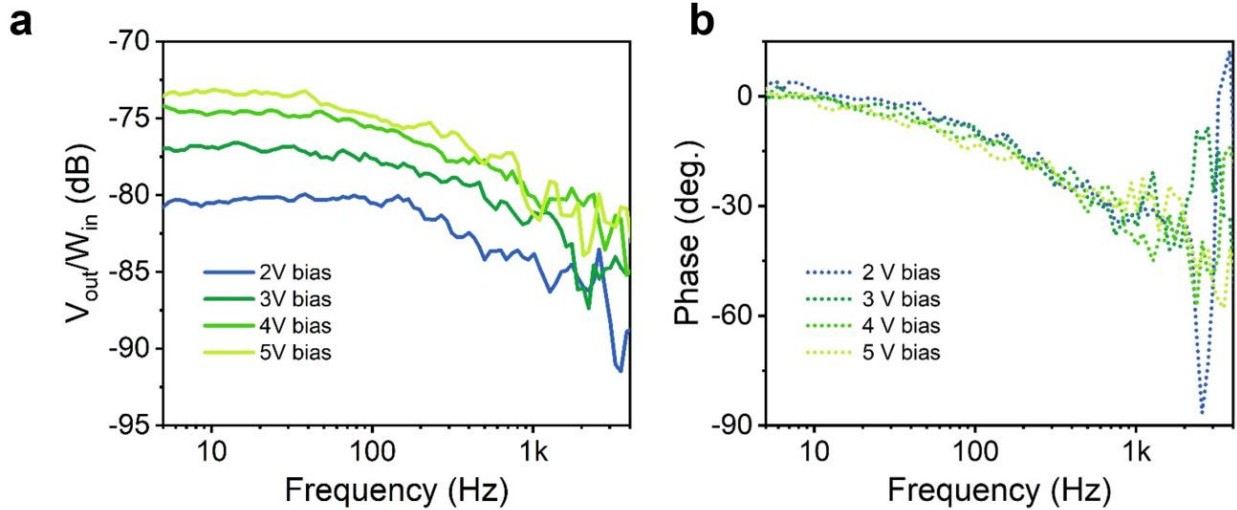

**Figure 5. Frequency-dependent photo-response.** AC photocurrent output (read as voltage after amplification, $V_{out}$) over the optical power input (a.u.) $W_{in}$ in dB versus frequency expressed in magnitudes **(a)** and phases **(b)**. Different bias voltages (blue-green color scales for magnitudes (solid lines) and phases (dotted lines)) are applied to the detector to test the frequency response. Corning glass substrates with gold contacts are used to reduce substrate capacity.

Finally, we exploit one of the most appealing features of NC films, i.e. the possibility to grow them at low temperature on a wide range of substrates by a versatile *in situ* reaction, to demonstrate potential application of the active material in a flexible UV sensor. Bottom metal contacts were prepared on PEN substrates (see Methods), later used for [AgSePh]$_\infty$ NC growth. Figure 6 shows the electrical characteristic of the detector tested against bending: during 6.4 mm bending radius (Figure 6c) and after ~2 mm bending radius (Figure 6d). The IV curves (Figure 6a) and the current response over intermittent illumination (Figure 6b) are basically unaltered upon bending. Such result demonstrates the appealing functionality of the [AgSePh]$_\infty$ NC film as a semiconductor compatible with flexible substrates and devices.

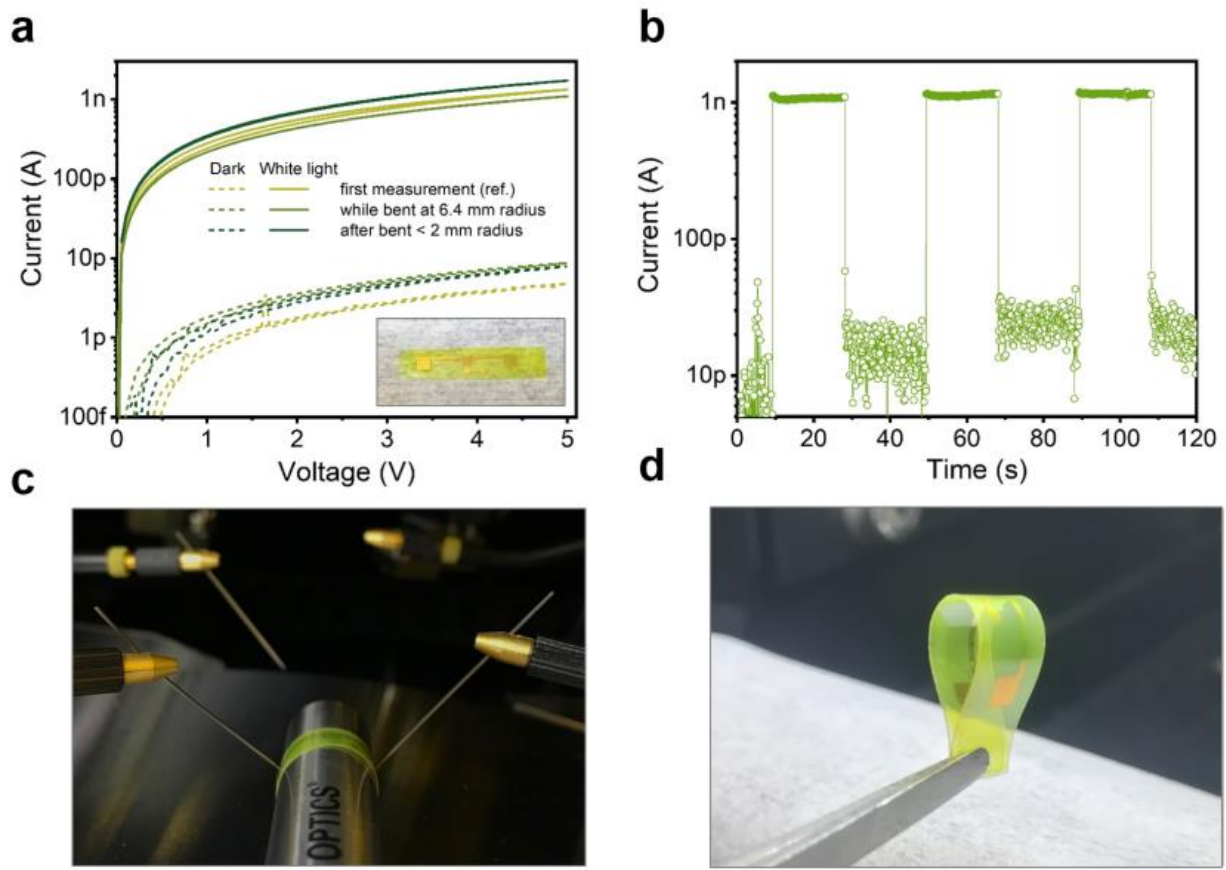

**Figure 6. Demonstration of working device on flexible substrate. a,** I-V characteristics in dark conditions and under white light illumination (~110 mW/cm$^2$) of a NCs film on polyethylene naphthalate (PEN) substrate (inset) contacted with thermal evaporated Ag contacts, with a channel length of 100 μm. **b,** Photo-current generated by switching the same white light ON and OFF over time, demonstrating measurement reproducibility. **c.** Photograph of the device working under bending conditions. **d,** Photograph of the bent test with ~2 mm bending radius.

CONCLUSION

We investigated for the first time the photo-electrical properties, and specifically the UV photo-response, of thin films of [AgSePh]$_\infty$, a multiple hybrid quantum well metal organic chalcogenide polymer. The continuous and smooth nanocrystal [AgSePh]$_\infty$ film was electrically contacted with high-work function electrodes to realize simple planar detectors. The devices show

< 20 pA dark-current levels, characterized by a high activation energy, and the possibility to extract a photocurrent under near UV illumination under ≤ 5 V applied bias, with a light-to-dark current ratio approaching 1000 under white light illumination. The good photo-responsivity is achieved thanks to a photoconductive gain mechanism confirmed at wavelengths shorted than 450 nm by an EQE exceeding 100% (270% at 370 nm). Under such working mechanism the detector operates with a cutoff frequency in the 300 – 500 Hz range. The lateral photo-conductor characteristics appears to be well within the range of the 2D perovskites thin film performance in terms of ON/OFF ratio, responsivity and cut-off frequency.[30–33] The NC film hereby presented is straightforward to grow on virtually any substrate and it can very well withstand bending, beyond $O_2$ and moisture, favorably comparing to other performing 2D materials exploited for UV detection, such as hybrid perovskites. This work indicates another appealing pathway for the emerging research on hybrid multiple quantum well structures based on metal chalcogenides polymers as promising tunable material platform for broadband optoelectronic applications.

MATERIALS AND METHODS

**Synthetic procedure:**

**Metal contacts patterning on substrates** was done by thermal evaporation of Cr/Au or Cr/Pt or Cr/Al (5 nm/ 40 nm) metal contacts after maskless optical lithography (SU8 photoresist on $SiO_2/Si^{++}$) to pattern interdigitated electrodes. The electrodes were designed to implement a device channel lengths of 5 or 10 or 20 μm with a fix channel width of 10 mm. Flexible devices were similarly prepared by Cr/Au evaporation (1.5 nm and 35 nm, respectively) on PEN substrates resulting in 75 μm channel length and 17.6 mm channel width.

**The [AgSePh]$_\infty$ nano-crystal (NC) film synthesis** started with 20 nm-thick silver films thermally evaporated on SiO$_2$/Si$^{++}$ substrates with pre-patterned metal contacts. The samples were then exposed to O$_2$ plasma to form AgO. Diener Electronic Femto Plasma asher was used. Molecular oxygen was injected in the chamber at pressure of 0.4 mbar and the plasma with nominal power of 100 W, regulated with forward (FW) and reflected power (BWD) in terms of percentages. The FW power plasma power was set to 10% with 1 min exposure time to obtain the best AgO film quality (while the BWD was kept at 0). Subsequently, the silver oxide films were exposed to a chemical vapor process. Benzeneselenol (97%, Sigma Aldrich) was inserted in a nitrogen glovebox in a Teflon-lined 22 ml vial next to the AgO covered substrate. The sealed vial was transferred in a pre-heated oven at 90°C. The reaction yielded [AgSePh]$_\infty$ after 4 hours; no electrical difference was actually observed for reactions exceeding 1 h; while for reactions longer than 24 h delamination affected some samples). All the samples were rinsed in acetone then isopropyl alcohol to remove the unreacted organo-chalcogen reagent and N$_2$ dried in a box, overnight.

**Morphological and structural characterization:**

**XRD** X-ray diffraction spectra were obtained with a BRUKER D8 ADVANCE diffractometer with Bragg-Brentano geometry equipped with a Cu K$\alpha_1$ ($\lambda = 1.5440$ Å) anode and operating at 40 kV and 40 mA. All the diffraction patterns were collected at room temperature over an angular range ($2\theta$) between 5° and 90° with step size 0.02° and 10 s acquisition time.

**SEM** Scanning electron microscopy images were collected with a JCM-6010LV, JEOL, with a secondary electron detector and at an accelerating voltage of 4 keV for the electron beam.

**UV-Vis** absorption spectra of [AgSePh]$_\infty$ NC films were derived by measuring transmission spectra on a PerkinElmer Lambda 1050 UV/Vis/NIR spectrometer.

**AFM** topography maps of NC films were collected in non-contact mode using an Agilent 5500 atomic force microscope operated in "acoustic mode". Specifications of the cantilever tip: L= 225 µm, W = 40 µm, T = 8.5 µm, $k_{el}$ = 48 Nm$^{-1}$, with a tip radius of curvature < 10 nm. The scan rate used was 0.75 inch/s with a resolution of 512x 512 pixels. Post processing was performed using Gwyddion software to level the data.

**Electrical characterization** was performed on a probe station connected to a semiconductor parameter analyzer (Agilent B1500A). Photocurrent measurements were carried out either under white light (microscope light, 115 mW/cm$^2$), or under monochromatic illumination supplied by several independent LED sources (except for the 254 nm wavelength illumination that was supplied by a bench-type UV lamp) calibrated at the same irradiance (~300 nW/cm$^2$). Frequency performances were measured using a custom setup that includes a Keysight ENA E5061B Vector Network Analyzer and an Agilent B2912A Sourcemeter. Assuming small voltage stimulus $V_{in}$, we linearized the LED response generating the optical signal $W_{in}$, obtaining $V_{out}/V_{in} \cong V_{out}/W_{in}$.

**XPS** analyses have been carried out using a Kratos Axis UltraDLD spectrometer. Data have been acquired using a monochromatic Al Kalpha source, operated at 20 mA and 15 kV. High resolution spectra have been acquired at pass energy of 10 eV, energy step of 0.1 eV and take-off angle of 0 degrees with respect to sample normal direction. Analysis area: 300 x 700 micron. Energy scale calibrated on C 1s at 284.8 eV.

**UPS** analyses were performed with the same setup, using a He I (21.22 eV) discharge lamp, on an area of 55 µm in diameter, at a pass energy of 5 eV and with a dwell time of 100 ms. The work function was measured from the threshold energy for the emission of secondary electrons during He I excitation. A −9.0 V bias was applied to the sample to precisely determine the low-kinetic-energy cutoff. The position of the VBM versus the vacuum level was estimated by measuring its distance from the Fermi level.

## ASSOCIATED CONTENT

The Supporting Information is available.

## AUTHOR INFORMATION

**Corresponding Authors:** lorenzo.maserati@iit.it, mario.caironi@iit.it

**Author Contributions:** L.M. and M.C. conceived the project. L.M. performed the synthesis, devices fabrication optical spectrometry, SEM, electrical and opto-electrical measurements. M.P. performed the XPS and UPS experiments. S.P. performed the X-ray diffraction. A.T. and S.P. performed the AFM experiment. A.P. and L.M. designed and carried out the frequency dependent sensor response experiment. B.P. performed the thin films thickness characterization. F.M. prepared the flexible substrates contacts.
L.M., D.N. and M.C wrote the manuscript.

**Funding Sources:** This work was in part financially supported by the European Research Council under the European Union's Horizon 2020 research and innovation program "HEROIC," grant agreement 638059.

## ACKNOWLEDGMENT

L.M. acknowledges the CNST technical staff for their support. This work was in part carried out at Polifab, the micro- and nanotechnology center of the Politecnico di Milano.

## REFERENCES

(1) Ariga, K.; Nishikawa, M.; Mori, T.; Takeya, J.; Shrestha, L. K.; Hill, J. P. Self-Assembly as a Key Player for Materials Nanoarchitectonics. *Sci. Technol. Adv. Mater.* **2019**, *20* (1), 51–95. https://doi.org/10.1080/14686996.2018.1553108.

(2) Li, H.; Eddaoudi, M.; O'Keeffe, M.; Yaghi, O. M. Design and Synthesis of an Exceptionally Stable and Highly Porous Metal-Organic Framework. *Nature* **1999**, *402* (6759), 276–279. https://doi.org/10.1038/46248.

(3) Givaja, G.; Amo-Ochoa, P.; Gómez-García, C. J.; Zamora, F. Electrical Conductive Coordination Polymers. *Chem. Soc. Rev.* **2012**, *41* (1), 115–147. https://doi.org/10.1039/C1CS15092H.

(4) Veselska, O.; Demessence, A. D10 Coinage Metal Organic Chalcogenolates: From Oligomers to Coordination Polymers. *Divers. Coord. Chem. Spec. Issue Honor Prof Pierre Braunstein - Part II* **2018**, *355*, 240–270. https://doi.org/10.1016/j.ccr.2017.08.014.

(5) Pedesseau, L.; Sapori, D.; Traore, B.; Robles, R.; Fang, H.-H.; Loi, M. A.; Tsai, H.; Nie, W.; Blancon, J.-C.; Neukirch, A.; Tretiak, S.; Mohite, A. D.; Katan, C.; Even, J.; Kepenekian, M. Advances and Promises of Layered Halide Hybrid Perovskite Semiconductors. *ACS Nano* **2016**, *10* (11), 9776–9786. https://doi.org/10.1021/acsnano.6b05944.

(6) Mao, L.; Stoumpos, C. C.; Kanatzidis, M. G. Two-Dimensional Hybrid Halide Perovskites: Principles and Promises. *J. Am. Chem. Soc.* **2019**, *141* (3), 1171–1190. https://doi.org/10.1021/jacs.8b10851.

(7) Smith, M. D.; Connor, B. A.; Karunadasa, H. I. Tuning the Luminescence of Layered Halide Perovskites. *Chem. Rev.* **2019**, *119* (5), 3104–3139. https://doi.org/10.1021/acs.chemrev.8b00477.

(8) Mitzi, D. B.; Feild, C. A.; Harrison, W. T. A.; Guloy, A. M. Conducting Tin Halides with a Layered Organic-Based Perovskite Structure. *Nature* **1994**, *369* (6480), 467–469. https://doi.org/10.1038/369467a0.

(9) Papavassiliou, G. C.; Koutselas, I. B.; Terzis, A.; Whangbo, M.-H. Structural and Electronic Properties of the Natural Quantum-Well System $(C_6H_5CH_2CH_2NH_3)_2SnI_4$. *Solid State Commun.* **1994**, *91* (9), 695–698. https://doi.org/10.1016/0038-1098(94)00435-8.

(10) Kitazawa, N. Excitons in Two-Dimensional Layered Perovskite Compounds: $(C_6H_5C_2H_4NH_3)_2Pb(Br,I)_4$ and $(C_6H_5C_2H_4NH_3)_2Pb(Cl,Br)_4$. *Mater. Sci. Eng. B* **1997**, *49* (3), 233–238. https://doi.org/10.1016/S0921-5107(97)00132-3.

(11) Yao, W.-T.; Yu, S.-H. Synthesis of Semiconducting Functional Materials in Solution: From II-VI Semiconductor to Inorganic–Organic Hybrid Semiconductor Nanomaterials. *Adv. Funct. Mater.* **2008**, *18* (21), 3357–3366. https://doi.org/10.1002/adfm.200800672.

(12) Zhang, Y.; Dalpian, G. M.; Fluegel, B.; Wei, S.-H.; Mascarenhas, A.; Huang, X.-Y.; Li, J.; Wang, L.-W. Novel Approach to Tuning the Physical Properties of Organic-Inorganic Hybrid Semiconductors. *Phys. Rev. Lett.* **2006**, *96* (2), 026405. https://doi.org/10.1103/PhysRevLett.96.026405.

(13) Stranks, S. D.; Snaith, H. J. Metal-Halide Perovskites for Photovoltaic and Light-Emitting Devices. *Nat. Nanotechnol.* **2015**, *10* (5), 391–402. https://doi.org/10.1038/nnano.2015.90.

(14) Ortiz-Cervantes, C.; Carmona-Monroy, P.; Solis-Ibarra, D. Two-Dimensional Halide Perovskites in Solar Cells: 2D or Not 2D? *ChemSusChem* **2019**, *12* (8), 1560–1575. https://doi.org/10.1002/cssc.201802992.

(15) Chen, Y.; Sun, Y.; Peng, J.; Tang, J.; Zheng, K.; Liang, Z. 2D Ruddlesden–Popper Perovskites for Optoelectronics. *Adv. Mater.* **2018**, *30* (2), 1703487. https://doi.org/10.1002/adma.201703487.

(16) Thrithamarassery Gangadharan, D.; Ma, D. Searching for Stability at Lower Dimensions: Current Trends and Future Prospects of Layered Perovskite Solar Cells. *Energy Environ. Sci.* **2019**, *12* (10), 2860–2889. https://doi.org/10.1039/C9EE01591D.

(17) Huang, X.; Li, J.; Fu, H. The First Covalent Organic−Inorganic Networks of Hybrid Chalcogenides: Structures That May Lead to a New Type of Quantum Wells. *J. Am. Chem. Soc.* **2000**, *122* (36), 8789–8790. https://doi.org/10.1021/ja002224n.

(18) Li, Y.; Jiang, X.; Fu, Z.; Huang, Q.; Wang, G.-E.; Deng, W.-H.; Wang, C.; Li, Z.; Yin, W.; Chen, B.; Xu, G. Coordination Assembly of 2D Ordered Organic Metal Chalcogenides with Widely Tunable Electronic Band Gaps. *Nat. Commun.* **2020**, *11* (1), 261. https://doi.org/10.1038/s41467-019-14136-8.

(19) Liu, W.; Lustig, W. P.; Li, J. Luminescent Inorganic-Organic Hybrid Semiconductor Materials for Energy-Saving Lighting Applications. *EnergyChem* **2019**, *1* (2), 100008. https://doi.org/10.1016/j.enchem.2019.100008.

(20) Maserati, L.; Refaely-Abramson, S.; Kastl, C.; Chen, C. T.; Borys, N.; Eisler, C. N.; Collins, M. S.; Smidt, T. E.; Barnard, E. S.; Strasbourg, M.; Shevitski, B.; Yao, K.; Schriber, E. A.; Hohman, J. N.; Schuck, P. J.; Aloni, S.; Neaton, J.; Schwartzberg, A. M. Anisotropic 2D Excitons Unveiled in Organic-Inorganic Quantum Wells. *Mater. Horiz.* **2020**. https://doi.org/10.1039/C9MH01917K.

(21) Sun, L.; Campbell, M. G.; Dincă, M. Electrically Conductive Porous Metal–Organic Frameworks. *Angew. Chem. Int. Ed.* **2016**, *55* (11), 3566–3579. https://doi.org/10.1002/anie.201506219.

(22) Cuthbert, H. L.; Wallbank, A. I.; Taylor, N. J.; Corrigan, J. F. Synthesis and Structural Characterization of [Cu20Se4(M3-SePh)12(PPh3)6] and [Ag(SePh)]∞. *Z. Für Anorg. Allg. Chem.* **2002**, *628* (11), 2483–2488.

(23) Mihailetchi, V. D.; Wildeman, J.; Blom, P. W. M. Space-Charge Limited Photocurrent. *Phys. Rev. Lett.* **2005**, *94* (12), 126602. https://doi.org/10.1103/PhysRevLett.94.126602.

(24) Natali, D.; Caironi, M. Organic Photodetectors. In *Photodetectors : materials, devices and applications*; Electronic and Optical Materials; 2015.

(25) L. Shi; S. Nihtianov. Comparative Study of Silicon-Based Ultraviolet Photodetectors. *IEEE Sens. J.* **2012**, *12* (7), 2453–2459. https://doi.org/10.1109/JSEN.2012.2192103.

(26) Caranzi, L.; Pace, G.; Sassi, M.; Beverina, L.; Caironi, M. Transparent and Highly Responsive Phototransistors Based on a Solution-Processed, Nanometers-Thick Active Layer, Embedding a High-Mobility Electron-Transporting Polymer and a Hole-Trapping Molecule. *ACS Appl. Mater. Interfaces* **2017**, *9* (34), 28785–28794. https://doi.org/10.1021/acsami.7b05259.

(27) Venugopalan, V.; Sorrentino, R.; Topolovsek, P.; Nava, D.; Neutzner, S.; Ferrari, G.; Petrozza, A.; Caironi, M. High-Detectivity Perovskite Light Detectors Printed in Air from Benign Solvents. *Chem* **2019**, *5* (4), 868–880. https://doi.org/10.1016/j.chempr.2019.01.007.

(28) Iacchetti, A.; Natali, D.; Binda, M.; Beverina, L.; Sampietro, M. Hopping Photoconductivity in an Exponential Density of States. *Appl. Phys. Lett.* **2012**, *101* (10), 103307. https://doi.org/10.1063/1.4750239.

(29) Fang, H.; Hu, W. Photogating in Low Dimensional Photodetectors. *Adv. Sci.* **2017**, *4* (12), 1700323. https://doi.org/10.1002/advs.201700323.

(30) Zhang, Y.; Liu, Y.; Xu, Z.; Ye, H.; Li, Q.; Hu, M.; Yang, Z.; Liu, S. (Frank). Two-Dimensional (PEA)2PbBr4 Perovskite Single Crystals for a High Performance UV-Detector. *J. Mater. Chem. C* **2019**, *7* (6), 1584–1591. https://doi.org/10.1039/C8TC06129G.

(31) Mei, F.; Sun, D.; Mei, S.; Feng, J.; Zhou, Y.; Xu, J.; Xiao, X. Recent Progress in Perovskite-Based Photodetectors: The Design of Materials and Structures. *Adv. Phys. X* **2019**, *4* (1), 1592709. https://doi.org/10.1080/23746149.2019.1592709.

(32) Wang, L.; Xue, Y.; Cui, M.; Huang, Y.; Xu, H.; Qin, C.; Yang, J.; Dai, H.; Yuan, M. A Chiral Reduced-Dimension Perovskite for an Efficient Flexible Circularly Polarized Light Photodetector. *Angew. Chem. Int. Ed.* **2020**, *59* (16), 6442–6450. https://doi.org/10.1002/anie.201915912.

(33) Tian, X.; Zhang, Y.; Zheng, R.; Wei, D.; Liu, J. Two-Dimensional Organic–Inorganic Hybrid Ruddlesden–Popper Perovskite Materials: Preparation, Enhanced Stability, and Applications in Photodetection. *Sustain. Energy Fuels* **2020**, *4* (5), 2087–2113. https://doi.org/10.1039/C9SE01181A.

SYNOPSIS

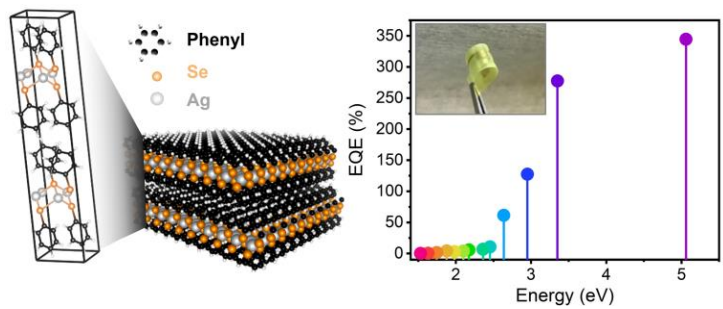